\documentclass{actasen}

\begin{document}

\setcounter{page}{0}

\Volume{2015}{0}


\runheading{Toshitaka TATSUMI }%

\title{Topological Aspects of the Inhomogeneous Chiral Phases and Implication on Compact Stars$^{\dag}$}

\footnotetext{$^{\dag}$ Partially supported by Grants-in-Aid for Scientific Research on Innovative Areas 
through No. 24105008 provided by MEXT.

\hspace*{5mm}$^{\bigtriangleup}$ tatsumi@ruby.scphys.kyoto-u.ac.jp\\

\noindent 0275-1062/01/\$-see front matter $\copyright$ 2011 Elsevier
Science B. V. All rights reserved. 

\noindent PII: }

\enauthor{Toshitaka TATSUMI$^{\bigtriangleup}$ }{Department of Physics, Kyoto University, Kyoto 606-8502, Japan}

\abstract{Recent studies of our group are reviewed about the inhomogeneous chiral phase. Appearance of the inhomogeneous chiral phase is a new paradigm in
the field of the QCD phase diagram. We consider the dual-chiral-density-wave in the presence of the magnetic field. Some topological effect manifests in the form 
of spectral asymmetry of the quark single-particle energy. }


\maketitle

\section{Introduction}
Nowadays implications of the QCD phase diagram on compact stars has become an intriguing subject in light of progress in observation.
Theoretically the QCD phase diagram has been extensively explored by the lattice QCD simulations or by using the effective models of QCD; chiral transition or deconfinement transition has been the main topic at finite temperature or density. 
Chiral transition has been usually studied by assuming the uniform and scalar quark-antiquark condensate. In 2005 we have suggested another possibility of the chiral transition at finite temperature or density, where the pseudoscalar condensate as well as the scalar condensate appears \rf{1}. Both condensates spatially modulated and we call it dual chiral density wave (DCDW).
Nowadays there are known various kinds of non-uniform condensate and they are described by the generalized order parameter, $M=\langle {\bar q}q\rangle+i\langle{\bar q}i\gamma_5\tau_3q\rangle\equiv \Delta({\bf r})\exp(i\theta({\bf r}))$, for the flavor $SU(2)$ case. Using the NJL model it has been shown that inhomogeneous chiral phases appear in the vicinity of the chiral transition. The critical point is then replaced by the Lifshitz point, where two uniform phases and one non-uniform phase meet \rf{2}.
 
Here we discuss the properties of the inhomogeneous chiral phase in the presence of the magnetic field.  We shall see that the phase degree of freedom becomes important once the magnetic field is taken into account.

\section{Inhomogeneous chiral phase in the magnetic field}
The phase transition in the magnetic field has been extensively studied and the enhancement of spontaneous symmetry breaking or magnetic catalysis has been found \rf{3-5}. 
The Dirac equation for the quark fields is also analytically solved in the DCDW phase. Frolov et al have shown that the DCDW phase is remarkably extended to the wide region once the magnetic field is applied \rf{6}. On the other hand, the real kink crystal (RKC) is also a popular configuration in the inhomogeneous chiral phase and has been well studied \rf{2}. 
In the recent work we have further studied this problem by taking into account the both RKC and DCDW configuration simultaneously, which we call the hybrid chiral condensate (HCC). We have found that there is competition between DCDW and RKC, and pure RKC phase never appears in the phase diagram. This is further discussed in the  Nishiyama's talk \rf{7}.

\section{Spectral asymmetry and Novel Lifshitz point}
One important feature is spectral asymmetry in the DCDW phase. Generally spectral asymmetry is defined as 
$
\eta_H=\lim_{s\rightarrow 0+}\sum_k{\rm sign}(\lambda_k)|\lambda_k|^{-s}
$
with $\lambda_k$ being the energy of the quark fields, which possesses a topological nature and 
 induces anomalous particle number \rf{8}.
We find that the energy spectrum of the quark fields are discretized to be the Landau levels in the presence of the magnetic field 
and the lowest Landau level exhibits spectral asymmetry \rf{9}.
Such asymmetry happens because of the spatially dependent phase of the condensate and the presence of the magnetic field, and closely related to chiral anomaly \rf{10}.

In the recent paper we have discussed an implication of spectral asymmetry on the lattice QCD simulation within the generalized Ginzburg-Landau theory \rf{9}. 
Consequently the Lifshitz point is shifted to the low $\mu$ and high $T$ region in the 
presence of the magnetic field, which should open a possibility to directly explore the phase transition to the DCDW phase by way of the lattice QCD simulation. 

\section{Phenomenological implications}

Here we briefly discuss two implications of DCDW on compact star phenomena.

\subsection{Spontaneous magnetization}
First, we'd like to suggest a possibility of spontaneous magnetization in the DCDW phase. The origin of the strong magnetic field in compact stars has been 
a longstanding issue since the first discovery of pulsars. Recent observation of magnetars has revived this problem. We can see that there is spontaneous 
magnetization in the DCDW phase due to the spectral asymmetry. Note that this phenomenon can be never revealed unless we examine the response to the external magnetic field.
 This subject will be discussed in detail by Yoshiike \rf{11}.   

\subsection{Fast cooling mechanism}
In a recent paper we have suggested a fast cooling mechanism due to DCDW \rf{12}. Considering the neutrino emission by the quark $\beta$ decay, it has been well known that 
it is strongly prohibited in compact stars due to the kinematic condition called the "triangle condition".
We have shown that 
the momentum conservation is modified at the weak-interaction vertex in the DCDW phase: the chiral condensate supplies or absorbs the extra momentum of the wave vector and 
thereby relaxes the kinematic condition. We have estimated the emissivity to find $10^{24-26}(T/10^6{\rm K})^6 {\rm ergs}\cdot{\rm cm}^{-3}\cdot{\rm s}^{-1}$, 
which is comparable with the one of the rapid cooling 
mechanism such as quark Durca or pion cooling. 
Some implication on the recent observation of Cas A should be interesting \rf{13}.

\section{Concluding remarks}
Some recent works about the DCDW phase have been reviewed. 
We have seen that the phase of the condensate plays an important role in the presence of the magnetic field. Spectral asymmetry has a deep meaning in this context; 
it remarkably extends the DCDW phase and gives rise to a novel Lifshitz point, which may be directly explored by the lattice QCD simulation. 
It then may be interesting and important to study the effect of the current quark mass for the Lifshitz point in a realistic situation.

\acknowledgements{The author likes to express hearty thanks to S. Karasawa, K. Nishiyama, R. Yoshiike and T. Muto for collaborations.}

\end{document}